\documentclass[aip,twocolumn,letterpaper]{revtex4}

\pdfoutput=1  

\usepackage{fullpage}

\usepackage{graphics}
\usepackage{epsfig}

\begin{document}
\title{The interaction of the ITER first wall with magnetic perturbations}
\author{Allen H. Boozer}
\affiliation{Columbia University, New York, NY  10027\\ ahb17@columbia.edu}

\begin{abstract}

Mitigation of the multiple risks associated with disruptions and runaway electrons in tokamaks involves competing demands.   Success requires that each risk be understood sufficiently that appropriate compromises can be made.  Here the focus is on the interaction of short timescale magnetic-perturbations with the structure in ITER that is closest to the plasma, blanket modules covered by separated beryllium tiles.  The effect of this tiled surface on the perturbations and on the forces on structures are subtle.  Indeterminacy can be introduced by tile-to-tile shorting.  A determinate subtlety is introduced because electrically separated tiles can act as a conducting surface for magnetic perturbations that have a normal component to the surface.  A practical method for including this determinate subtlety into plasma simulations is developed.  The shorter the timescales and the greater the localization, particularly in the toroidal direction, the more important the magnetic effects of the tiles become.

\end{abstract}

\date{\today} 
\maketitle

\begin{figure}
\centerline{ \includegraphics[width=2.0in]{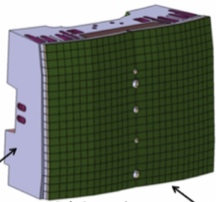} }
\caption{One of the 440 ITER blanket modules is illustrated.  Each is 1~m by 1.5~m by 0.5~m thick and contains approximately 4.6 tons of stainless steel.  The plasma side of the modules is covered by beryllium tiles, which are 0.8~cm thick and 5~cm wide.  The tiles are supported on stainless steel fingers, which extend in the toroidal direction.  Directly behind the beryllium tiles, each finger has a layer of copper in which water-carrying tubes are embedded.  The volume of copper is approximately twice that of the beryllium, and the copper has approximately twice the electrical conductivity. This figure is reproduced with permission from Nucl. Fusion \textbf{54}, 033004 (2014). Copyright 2014, Institute of Physics Publishing, where it is part of Figure 2. }
\label{fig:wall_f}
\end{figure}

\section{Introduction}

To ensure ITER provides as much information as possible, empirical, computational, and theoretical studies are required on disruption and runaway issues.  This paper is focused on issues associated with currents and magnetic forces during disruptions \cite{Disruption-scenarios:2007,ITER-disruptions:Lehnen,Lehnen:2020}, particularly due rapid transient events, $\lesssim50~$ms, which challenge the short-time limit of the ITER design.

There is a wide range views of the importance of disruptions and runaway electrons to the minimization of the risk, time, and cost of developing fusion power plants.  Some view it as slight---even addressable on existing tokamaks.  In 2019, Hawryluk and Zohm \cite{Physics Today:2019} expressed a common view that may sound consistent with retiring the risks using existing tokamaks: ``\emph{Touching an operational boundary can lead to the occurrence of MHD instabilities that can terminate the plasma discharge. In present experiments, the worst plasma confinement disruption occurs when a substantial plasma current of the order of 1~MA is lost rapidly on a time scale of several milliseconds. In ITER, 15~MA of plasma current could be rapidly terminated over a somewhat longer time. A sudden termination of the plasma discharge generates large toroidal electric fields that can drive energetic runaway electron currents, transmit heat fluxes to the plasma-facing components, and apply large forces to the vacuum vessel and its components. Mitigation systems are being implemented to ameliorate the disruptions if the operating boundaries are exceeded, and active control systems will be used to ensure that the plasma operates within safe boundaries.}"  Others \cite{CO2-Stellarator} view the importance as sufficiently great that a prudent program would more carefully consider the benefits of fusion systems that avoid these risks. 

However one assesses the risks of disruptions and runaway electrons, program plans should be chosen to be consistent with the science.  The science cannot be chosen to be consistent with the program plans.

The most concerning issue may be the formation of beams of runaway electrons (RE) on high-current tokamaks.  The fundamental problem is that a change in the poloidal magnetic flux outside of the region where an electron is confined provides an accelerating voltage, and high-energy electrons, $\sim10~$MeV, can directly scatter a cold electron to a sufficiently high energy that the acceleration is greater than the collisional drag.  In hydrogenic plasmas, each megaampere drop in the plasma current amplifies the number of accelerating electrons by a factor of approximately ten.  It is concerning that the danger has been known for more than two decades \cite{{R-P-runaway1997}}, yet  ``\emph{With ITER construction in progress, reliable means of RE mitigation are yet to be developed}."   This was noted by Breizman, Aleynikov, Hollmann, and  Lehnen  \cite{Breizman:2019} in the last paragraph of their 2019 review.   In Section 11.3, they said the required runaway current to produce melting if the runaway current struck the first wall was about 300 kA if concentrated or 1.9 MA if spread out.  

As discussed in 2020 by Vallhagen, Embreus, Pusztai, Hesslow, and F\"ul\"op \cite{Fulop:runaway2020}, the runaway issues in tokamaks become more difficult during DT operations due to high-energy electron production by tritium decay and Compton scattering by gamma rays from irradiated walls.   In non-nuclear experiments, the only seed for runaway amplification is the remaining high-energy electrons from the pre-disruption plasma, and they are all lost when the magnetic surfaces are broken for at least 15~ms during the thermal quench.  Impurities injected to mitigate the force and power loading associated with disruptions can greatly enhance the amplification in the number of runaway electrons per Ampere drop in the plasma current \cite{Fulop:runaway2020}.  

The runaway danger is a risk-probability balance, much as choosing the speed to drive on an icy road.   Once a tokamak is operating at a current above a certain value, the possibility of severe damage from relativistic electrons is present.  For success, ITER must achieve its mission with less than approximately one major runaway incident per thousand shots.   The duty-factor and repair-cost requirements of a practical power plant makes the constraints on runaway incidents even greater than in ITER.   ITER will provide no data on runaway issues in DT plasmas for at least fifteen years, and presumably a power plant must be operated for some years to demonstrate that an ITER-like power plant has overcome the issues of disruptions and runaway to the point that fusion can be extensively deployed as an energy source.  
 
 Turco and coauthors have reported \cite{Turco:2020}  that DIII-D has had no disruptions when operating at $q_{95}$ greater than five.  If by a disruption one means the sudden loss of a large fraction of the magnetic surfaces, this is not an unexpected result.  Unfortunately, if the plasma cools for any reason: (a) Axisymmetric control can be lost if the cooling is faster than the time scale for changing the vertical field.  (b) The runaway electron problem can be made worse, because of the confinement of relativistic electrons by the magnetic surfaces, unless the cooling is slower than the time scale for the quenching of the plasma current.

Disruptions and runaways are actually three coupled issues---a solution to one tends to complicate the solution to the other two: (1) loss of position control of the plasma, (2) excessive power deposition, particularly on the divertor, and (3) wall melting due to runaway electrons.  Despite conflicting requirements, all three risks must be retired before tokamak power plants can be deployed.  All three risks must be carefully studied to obtain an adequate compromise among them.  For example, the rapid quench of the plasma temperature, which occurs during disruptions, could cause unacceptable melting if the plasma energy were carried to the divertor by plasma transport or deposited on a localized wall region.  A mitigation strategy is to use impurity radiation to spread the heat uniformly on the walls, but this exacerbates the runaway electron problem.  

The focus of this paper is on the first issue, the loss of position control and magnetic effects when time scales are short $\lesssim50~$ms. 

 The distance of the superconducting coils that produce poloidal field are far from the ITER plasma and respond so slowly.   Even with the aid of two resistive coils. vertical position control is easily lost \cite{Gribov:Vertical2015,Lukash:vertical2013} when the plasma pressure and internal inductance undergo substantial changes that are fast compared to the response time of the vertical field from the superconducting coils. When position control is lost, the ITER plasma will drift vertically until it strikes the chamber wall, which can deposit the runaway electrons in the plasma on the walls and produce large forces through the effects of induced and halo currents.  Halo currents flow for part of their path through the wall and for part through the halo plasma, which by definition is the part of the plasma that has a short connection length with the wall along the magnetic field lines.  
 
 Although rapid plasma cooling exacerbates the runaway problem, the planned method for avoiding the effects of vertical displacements is a rapid plasma cooling, to approximately 20~eV, to make the current quench faster than 150~ms.  This ensures that both the normal magnetic field to the walls and the poloidal flux that can enter the chamber undergo little change during the current quench.  A fast current quench compared to the resistive time scale of the chamber is fundamentally different than the situation studied in existing tokamaks such as JET and DIII-D.  The 150~ms limit was thought to ensure that the loss of axisymmetric position control could not result in the plasma striking the wall while carrying a substantial current.  However,  Kiramov and Breizman \cite{Breizman:2018} showed in 2018, that a tokamak plasma enclosed by a perfectly conducting wall can move vertically and strike the walls as its current decays.  Boozer \cite{Ideal VDE}  showed that this motion can even become unstable.  Both effects were confirmed in M3D-C1 axisymmetric calculations by Clauser and Jardin \cite{Clauser:2020}.  The implications for ITER require more simulations since these  effects are profile and shape dependent.  Even when the current quenches before the plasma strikes the wall, currents are induced in the wall that can produce excessive forces on the individual blanket modules, Figure \ref{fig:wall_f}, rather than having the forces transferred to the vacuum vessel.  This can place a limit on the fastest current quench   \cite{Disruption-scenarios:2007,ITER-disruptions:Lehnen,Lehnen:2020}, which is generally given as 50~ms. 

The 50~ms limit on the time scale is  subtle.  A time scale faster than the penetration time through a conductor does not increase the force integrated over the volume of the conductor---just the spatial distribution is changed, Appendix \ref{sec:approx}.  The depth $d=\sqrt{(4\eta/\mu_0)\tau}$ that a magnetic field can penetrate into a block of stainless steel when $\tau=50$~ms is $\approx33$~cm.  The blanket modules \cite{ITER:blanket2014} are approximately 50~cm thick but do have internal structures to speed the penetration.   The increase in overall force on the blanket modules for time scales shorter than 50~ms is unclear.

\begin{figure}
\centerline{ \includegraphics[width=2.0in]{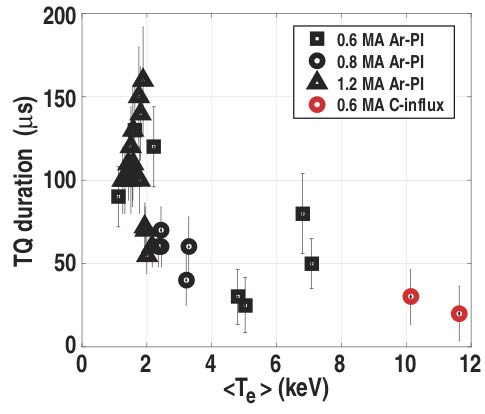} }
\caption{Extremely fast quenches of the electron temperature were observed \cite{Paz-Soldan:2020} during DIII-D disruptions.  Fast quenches of the electron temperature are expected \cite{Boozer:j flattening} when the magnetic surfaces are destroyed. The scaling of the time required for the electrons to cool by a collisionless streaming along chaotic magnetic field lines is $R_0/\sqrt{T_e}$, where $R_0$ is the major radius and $T_e$ is the electron temperature.  The major radius of ITER is larger than that in DIII-D by 6.2~m/1.67~m$\approx3.7$.   The constraint on the electron temperature for the mean-free-path to be long compared to the distance along a chaotic magnetic field line from the center to the edge scales as $\sqrt{n_eR_0 }$, where $n_e$ is the number density of electrons.  This figure is reproduced with permission from Nucl. Fusion \textbf{60}, 056020 (2020). Copyright 2020, Institute of Physics Publishing. }
\label{fig:TQ}
\end{figure}

If a large violation of the 50~ms limit on magnetic field changes would produce unacceptable machine damage, it may be necessary to address the force issue by reducing the maximum magnetic field rather than by trying to limit the speed with which magnetic field changes can be made.  A 50~ms limit has implications for ITER operations:

\begin{enumerate}
\item To ensure the current quench is faster than 150~ms, the resistivity temperature must be approximately 20~eV.  The resistivity temperature is the temperature a hydrogenic plasma would have to have to reproduce the rate of current decay.  The resistivity depends on $T_e^{3/2}$, so a reduction of the resistivity temperature by a further factor of two, would reduce the current-quench time to 53~ms.   The resistivity temperature can be changed not only by reducing the overall temperature but also by breaking the magnetic surfaces.  With broken surfaces, the current profile broadens all the way to the wall and the resistivity temperature can be determined by a coldest part of the halo plasma \cite{Boozer:helicity}.  The short magnetic connection length of the halo plasma to the walls causes a rapid flow of thermal energy, which keeps that plasma relatively cold.

\item Disruption phenomena can produce much faster changes in the magnetic field than 50~ms.  Changes in the electron temperature as fast as 0.025~ms were seen on DIII-D \cite{Paz-Soldan:2020},  Figure \ref{fig:TQ}.  Such fast time scales are expected when a thermal quench is initiated by the breakup of the magnetic surfaces in a multi-kilovolt plasma,  \cite{Boozer:j flattening}.  As discussed in Section \ref{sec:off normal}, rapid changes in the electron pressure give rapid changes in the magnetic field at the walls.   The resulting forces on the blanket modules are sensitive to the direction in which currents flow, which can be subtle.  A given current flowing in the poloidal direction produces an order of magnitude larger force than it would flowing in the toroidal direction, for the toroidal magnetic field is an order of magnitude larger than the poloidal field.

\item Possibly the most important consequence of having a 50~ms limitation is that it excludes interesting methods for preventing unacceptable machine damage from relativistic electrons.  In practice, this limits the exploration of such methods.  An example is filling the plasma with particles of tungsten when the number of relativistic electrons exceeds some level.  It is thought that this would reduce the plasma temperature and make the current quench far faster than 50~ms.  With the present understanding of the relative dangers, the exclusion from consideration of methods of removing large currents of relativistic electrons because of their associated fast current quench appears odd.

\end{enumerate}

The focus of this paper is physics that can change the path of wall currents during fast disruption phenomena.  When time scales become short, the properties of the conductors closest to the plasma become of upmost importance.  These are the tiles and the stainless steel fingers covered with copper as described in the caption to Figure \ref{fig:wall_f}.  When the tile/finger structure is viewed as continuous, the  formula $d=\sqrt{(4\eta/\mu_0)\tau}$ implies that for times less than 3~ms the magnetic field cannot penetrate that structure.   For longer times, the electric field $\vec{E}$ would be essentially constant across the tile/finger structure with the current density $\vec{j}=\vec{E}/\eta$.  Force-producing currents could continue to flow in a continuous tile/finger structure over a far longer time than 3~ms, Appendix \ref{sec:approx}.

The tile/finger structure is, however, not continuous.  The poloidal gaps between the fingers can block poloidal currents, which are the more dangerous since they interact with the toroidal magnetic field, and the toroidal gaps between the tiles can block toroidal currents.  This paper discusses the physics of these complicated structures and a tractable mathematical method for addressing them.


A clear distinction must be drawn among the three types of currents that can flow in a continuous toroidal structure surrounding the plasma: (1) net currents, (2) dipolar currents, and (3) halo currents.  These three types of current will be discussed in more detail in Section \ref{sec:cp}, but an understanding of the distinction is so essential that a brief description is given here.  

Two net currents, $I_s$ and $G_s$, are spatial constants within a thin conducting structure called a shell.  The net toroidal current $I_s$ controls the poloidal magnetic flux exterior to a toroidal shell.  When the structure is a perfect conductor $I_s$ must take whatever value is required to maintain a fixed poloidal flux outside the shell---the magnetic flux through the hole of torus.  The net poloidal current $G_s$ controls the toroidal magnetic flux enclosed by the shell.  When the structure is a perfect conductor $G_s$ must take whatever value is required to maintain a fixed enclosed toroidal magnetic flux.  The definition of the two net currents is made unique by choosing the dipolar currents to produce all of the magnetic field perpendicular to the structure.

The dipolar currents control the distribution of magnetic field normal to the shell and produce no poloidal or toroidal magnetic flux.  The name comes from the fact that any distribution of magnetic field normal to the shell could be produced by a collection of magnetic dipoles embedded in the shell.  A set of electrically isolated tiles covering the structure can carry a dipolar current but can carry neither a net nor a halo current.  Both the net and the dipolar currents can be represented in a thin conducting structure by a current potential $\kappa(\theta,\varphi)$, which depends on the poloidal $\theta$ and the toroidal  $\varphi$ angles, Section \ref{sec:cp}.

Halo currents, Section \ref{sec:halo}, flow for only part of their path through the conducting shell; the rest of their path is through the plasma.  They are defined within the conducting shell so they produce neither a normal magnetic field nor change the poloidal or toroidal magnetic fluxes.


Large forces in unexpected directions could be exerted on the tiles and blanket modules for two reasons:  (1) The channels between the tiles could be shorted by plasma entering or being formed in the channels or by localized melting bringing the beryllium tiles into electrical contact.  (2) The beryllium tiles, over sufficiently short times $<<20~$ms, act as a collection of individual magnetic dipoles, which can make the response of the first wall equivalent to that of a conductor for changes in the normal magnetic field to the first wall.  The dipolar effect is only important in the absence of large-scale shorting, although the effect is always in principle present.

The issue of whether the channels between tiles will be shorted is subtle, and the evolution of the shorting as the machine ages is difficult to predict.  Current flows between tiles may have been seen on the Compass tokamak in Prague \cite{Compass:2020}.  Tile melting was observed on JET with an ITER-like wall  \cite{JET-melt}.  These observations naturally raise concerns about shorting and its indeterminacy.  

The complicated current paths that can be caused by partial shorting can produce larger forces than complete shorting. For example, what would naturally be toroidal currents can be forced to have a path that crosses the much stronger toroidal magnetic field.  Partial shorting can also force the current to flow in a narrow channel, which intensifies the local force.   

The emphasis of the paper will be on the dipolar cause of large forces because of (1) its certainty, (2) the lower level of general familiarity, and (3) absence of tractable alternatives for making calculations.  The primary result is a simplified treatment: a pair of jump conditions that are valid when tiling is approximated as lying in a thin shell.   A jump condition is denoted by $[\cdots]$ and means the change a quantity between the two sides of the shell. 
\begin{itemize}
\item The normal-field, $B_\bot\equiv\hat{n}\cdot\vec{B}$ with $\hat{n}$ the unit normal to the shell, obeys the jump condition
\begin{equation}
\Big[ B_\bot \Big] =0.
\end{equation}

\item The tangential-field, $\vec{B}_{tan} \equiv \vec{B}-B_\bot\hat{n}$, obeys the jump condition
\begin{equation}
\Big[ \vec{B}_{tan} \Big] =\vec{\nabla}(s_tB_\bot).
\end{equation}

\end{itemize}
The jump conditions allow the dipolar effect to be added to calculations made using the well known two and three dimensional magnetohydrodynamics (MHD) codes. The magnitude of the effect due to the tiles is quantified by a length $s_t=d_tw_t/\delta_e$, where $d_t$ is the thickness of the tiles, $w_t$ the width of square tiles, and $\delta_e$ the effective separation between tiles.  The effective separation would be the actual separation if the tiles were perfect conductors, but $\delta_e$ is broadened for the tiles in ITER by resistive penetration.  Section \ref{sec:tile resis} shows that  $w_t/\delta_e = \sqrt{\tau_t/\tau_g}$, where the resistive time scale of the tiles is $\tau_t \equiv \mu_0 w_t^2/4\eta\approx 20~$ms.  The normal field is assumed to grow as $B_\bot\propto e^{t/\tau_g}$.  

For a given set of tiles and supporting structure, the length scale $s_t$ depends on the position on the first wall and on the frequency with which the magnetic field changes.  An empirical $s_t$ can be obtained using a magnetic probe and should be used in studies of the effects of the plasma on wall forces.  Theory is important for suggesting the form of the interaction and for estimating $s_t$, but the theory is not sufficiently precise to eliminate the need for determining $s_t$ empirically.

The forces due to the dipolar effect are sensitive to both the wavelength, particularly in the toroidal direction, and the growth rate $\tau_g$ of $B_\bot$.  The shorter the wavelength and $\tau_g$ the more accurately the tiled wall approximates a perfect conductor.  A tiled wall has no magnetic interaction when $B_\bot=0$.   

It is interesting to note that the role of the normal and the tangential field is reversed in the jump conditions for a conducting shell from those in which the shell consists of a linear magnetic material with $\mu/\mu_0>>1$, Appendix \ref{sec:magnetization}.

Section \ref{sec:cp} explains how the induced current flowing in a thin shell can be represented by a current potential $\kappa$.  The current potential has two parts:  (1) The single-valued part, which is associated with the normal magnetic field to the shell, has different forms for tiles and for continuous conductors; forms for both cases are derived in  Section \ref{sec:svcp}.  (2) The non-single-valued part of the current potential, which is associated with the poloidal and the toroidal  and the poloidal magnetic fluxes,  represents the two net currents $I_s$ and $G_s$ as discussed in Section \ref{sec:nsvcp}.  Halo currents produce a third pattern of current in a shell, and this is derived in Section \ref{sec:halo}.

 Section \ref{sec:off normal} discusses the implications of the results of this paper on the study of off-normal events in tokamaks.
 
 Appendix \ref{sec:approx} derives necessary conditions for the validity of the current-potential approximation.   For comparisons, Appendix \ref{sec:magnetization} derives the jump conditions for a thin shell that contains magnetic rather than electrically-conducting material. 


\section{Representation of current in a shell \label{sec:cp}}

A conducting structure of depth $d_s$ can be viewed as a thin shell when $d_sk<<1$, where $k$ is the wave number of the external magnetic field, and the electric field can be taken to be constant across its depth $d_s$ when $d_s<<\sqrt{\eta \tau_g/\mu_0}$, where $\tau_g$ is the time scale for changes in the external magnetic field.  The condition for the constancy of the electric field across the shell is the same as for the jump in normal component of the magnetic field to be zero, Appendix \ref{sec:approx}.

The forces and the external effects of a current $\vec{j}$ flowing in a thin toroidal shell can be represented by a current potential $\kappa(\theta,\varphi)$, where $\theta$ is an arbitrary poloidal and $\varphi$ an arbitrary toroidal angle.   The concept of a current potential was introduced into toroidal plasma physics in 1987 by Peter Merkel \cite{Merkel:1987}.  Current potentials have the form
\begin{eqnarray}
\kappa(\theta,\varphi) &=& \frac{G_s}{2\pi}\varphi - \frac{I_s}{2\pi}\theta+\tilde{\kappa}(\theta,\varphi) \mbox{    with   }\\
\vec{j}&=&\vec{\nabla} \times\Big( \kappa(\theta,\varphi)\delta(r-a) \hat{n}\Big). \label{j-eq}
\end{eqnarray}
$G_s$ is a constant, the number of Amperes of poloidal current, which produces the net toroidal field, $I_s$ is a constant, the number of Amperes of toroidal current, which produces the net poloidal field, and $\hat{n}$ is the unit normal to the surface.  The unit normal $\hat{n}$ is curl free, Equation (\ref{grad r}).  The $(r,\theta,\varphi)$ coordinate system that is used to describe the region near the shell is defined in Equation (\ref{coord-system}), and $r=a$ implies lying on the shell itself.   Magnetic dipoles oriented normal to the shell can represent the single-valued part of the current potential $\tilde{\kappa}(\theta,\varphi)$. 

Equation (\ref{j-eq}) assumes the current in the shell is divergence free, but this assumption is incorrect in the presence of halo currents, which have part of their path in the plasma.  Halo currents are discussed in Section \ref{sec:halo}.

 When $\theta$ and $\varphi$ are arbitrary, an additive part of $\tilde{\kappa}$ must be chosen to be proportional to $G_s$ and $I_s$ to ensure these currents produce no normal magnetic field.   New angles can always be defined by $\theta_n=\theta+\lambda(\theta,\varphi)$ and $\varphi_n=\theta+\omega(\theta,\varphi)$, where $\lambda$ and $\omega$ are arbitrary periodic functions.  The part of $\tilde{\kappa}$ that ensures that $G_s$ and $I_s$ produce no normal magnetic field can be absorbed into $\lambda$ and $\omega$.

The practical implications differ between the single-valued and the non-single-valued parts of the current potential.  The single-valued part is considered in Section \ref{sec:svcp} and the non-single-valued part in Section \ref{sec:nsvcp}.  

Single-valued current potentials are of two types depending on whether they represent: (1) electrically isolated tiles, Section \ref{sec:tcp}, or (2) a continuous conductor, Section \ref{sec:ccp}.  The two have distinct behaviors, but both are modified by a nearby conducting wall in a way that has a general form when the separation is a constant distance $\Delta$, Section \ref{sec:shielding}.   Section \ref{sec:ncw} derives the effect of a nearby conductor on the force on separated tiles.  

 The poloidal current $G_s$, which modifies the toroidal magnetic field is considered in Section \ref{sec:tfr}, and the toroidal current $I_s$, which modifies the poloidal field is considered in Section \ref{sec:pfr}. 


\subsection{Mathematical properties}

The general mathematical properties of the current potential and its existence were derived by Boozer in 2015, Section 2 of \cite{Boozer:3D}.   A toroidal surface is described by the function $\vec{x}_s(\theta,\varphi)$, which means the three Cartesian coordinates are given as functions of $\theta$, an arbitrary poloidal, and $\varphi$, an arbitrary toroidal angle.   Points in space near that surface can be represented by
\begin{eqnarray}
\vec{x}(r,\theta,\varphi) &=& \vec{x}_s(\theta,\varphi) + (r-a)\hat{n}.  \label{coord-system}\\
\hat{n} &=& \frac{ \frac{\partial \vec{x}_s}{\partial\theta}\times \frac{\partial \vec{x}_s}{\partial\varphi}}{\left|  \frac{\partial \vec{x}_s}{\partial\theta}\times \frac{\partial \vec{x}_s}{\partial\varphi}  \right|} \nonumber\\
&=& \vec{\nabla}r. \label{grad r}
\end{eqnarray}
The relation between the unit normal and the gradient of the radial coordinate, $\hat{n}=\vec{\nabla}r$, follows from the theory of general three-dimensional coordinate systems.  The general theory of  three-dimensional coordinate systems is derived in less than two pages in the Appendix to \cite{Boozer:3D}.  

Ignoring halo currents, the current density in the shell $\vec{j}$ must be divergence free and can only flow in the two tangential directions.  The implication is that it can be represented as the curl of a vector in the third direction, $\hat{n}$, Equation (\ref{j-eq}).

The single-valued part of the current potential $\tilde{\kappa}$ is the magnetic moment, $\vec{m}$ per unit area as proven in Section 2.2.1 of \cite{Boozer:3D},
\begin{eqnarray}
\vec{m}&\equiv&\frac{1}{2}\int (\vec{x}\times\vec{j}) d^3x \nonumber\\
& =& \hat{n} \int \tilde{\kappa} da.
\end{eqnarray}

The current density integrated across the shell is the surface current and has the units of Amperes per meter, 
\begin{eqnarray}
\vec{\mathcal{I}}&\equiv& \int \vec{j} dr \\
&=& \vec{\nabla}\kappa\times\hat{n}. \label{I-kappa}
\end{eqnarray}


\subsection{Effects on the external magnetic field}

The effect of the current potential on the magnetic field outside the shell can be studied using the relation $\vec{B} =\vec{\nabla}\phi + \mu_0 \kappa(\theta,\varphi)\delta(r-a) \hat{n}$, which follows from the observation that the curl of this expression is $\vec{\nabla}\times\vec{B}-\mu_0\vec{j}=0$.  The shell is assumed sufficiently thin that $\vec{\nabla}\cdot\vec{B}=0$ implies the normal component of the magnetic field, $\hat{n}\cdot\vec{B}$ cannot change across the shell.  Three jump conditions are implied:
\begin{itemize}

\item The jump in the magnetic scalar potential
\begin{eqnarray}
\Big[\phi \Big] &\equiv& \phi_+ - \phi_- \\
&=& - \mu_0\kappa, \label{phi-jump}
\end{eqnarray}
where  $\phi_+(\theta,\varphi)$ is evaluated at $r$ infinitesimally larger than $a$ and $\phi_-$ at $r$ infinitesimally smaller than $a$. 

\item The jump in the normal magnetic field
\begin{eqnarray}
\Big[ B_\bot  \Big] &=& \left[ \frac{\partial\phi}{\partial r} \right] \label{B-perp jump} \\
&=&0. \label{B-perp jump}
\end{eqnarray}
where $B_\bot \equiv \hat{n}\cdot\vec{B}$ is the part of the field that is normal to the shell.

\item The jump in the tangential magnetic field
\begin{eqnarray}
\Big[\vec{B}_{tan} \Big]&=& -\mu_0\vec{\nabla}\kappa, \label{B-tan jump}
\end{eqnarray}
 where $\vec{B}_{tan}\equiv\vec{B}-B_\bot\hat{n}$ is the part of the magnetic field that is tangential to the shell.  The theory of general coordinates can be used to prove that $\hat{n}\cdot\vec{\nabla}\kappa(\theta,\varphi)=0$.

\end{itemize}


\subsection{The magnetic force on the shell}

The Lorentz force density $\vec{f}_L =\vec{j}\times\vec{B}$ integrated across the shell, $\vec{F}_L=\int \vec{f}_L dr$, is the force per unit area applied by the magnetic field to the shell.  The normal component, $\hat{n}\cdot\vec{F}_L$, is equivalent to the magnetic pressure applied to the shell, but there is also a Lorentz force tangential to the shell.  Assuming the current density is constant across the shell, so $\vec{j}=\vec{\mathcal{I}}/d_s$ within the shell, where $d_s$ is its thickness
\begin{eqnarray}
\vec{F}_L &=& \vec{\mathcal{I}} \times \int \vec{B}\frac{dr}{d_s} \\
&=& \vec{\mathcal{I}} \times \left(\vec{B}_- +\frac{1}{2} \Big[ \vec{B} \Big]\right) \label{mid-eq} \\
&=& \left(\frac{\vec{B}_+ +\vec{B}_-}{2} \cdot\vec{\nabla}\kappa \right)\hat{n} -\left(\vec{B}\cdot\hat{n}\right)\vec{\nabla}\kappa. \label{shell force}
\end{eqnarray}
Equation (\ref{mid-eq}) gives the expression for the integral $(\int \vec{B} dr)/d_s$ when the current density is assumed to be a spatial constant across the thin shell.  Tangential magnetic fields produce forces that are normal to the first wall.  Normal magnetic fields produce forces that are both normal and tangential to the wall. 

Equation (\ref{shell force}) implies that current potentials that change over a short distance in the toroidal direction produce particularly large forces because of their interaction with the toroidal magnetic field.  The toroidal field is an order of magnitude stronger than the poloidal in tokamaks.


\subsection{Generalization for halo currents \label{sec:halo}}

The general shell-current is not divergence free because a halo current $J_h(\theta,\varphi)$ can flow between the plasma and  the shell.   For part of its path, this current flows through the halo plasma, which is the part of the plasma that is outside the closed magnetic surfaces.  An equal number of Amperes of halo current must flow from the shell into the plasma as flows from the plasma into the shell.  As will be shown, the implications are that $\oint J_h d\theta d\varphi=0$ and $J_h$ is a periodic function of $\theta$ and $\varphi$.  This implies $J_h$ is distinct from the two net currents $G_s,I_s$, but $J_h$ has a part that contributes to the dipolar current, which is represented by $\tilde{\kappa}$, as well as a part that does not. 

The general expression for the current in a thin shell that represents a wall is 
\begin{eqnarray}
\vec{j}&=&  \vec{\mathcal{K}}(\theta,\varphi)\delta(r-a)\times \hat{n} \nonumber\\ &&\hspace{0.2in} + J_h(\theta,\varphi) H(a-r) \vec{\nabla}\theta\times\vec{\nabla}\varphi; \\
\vec{\nabla}\cdot\vec{j} &=&\Big\{ \hat{n}\times (\vec{\nabla}\times \vec{\mathcal{K})}  - \frac{J_h}{\mathcal{J}} \Big\}\delta(r-a), \mbox{   where   } \hspace{0.2in}\\
\vec{\mathcal{K}} &=& \left(\frac{ \partial \kappa}{\partial\theta} + \tilde{\mathcal{K}}_\theta \right)\vec{\nabla}\theta + \left(\frac{ \partial \kappa}{\partial\varphi} + \tilde{\mathcal{K}}_\varphi \right)\vec{\nabla}\varphi;  \\
\mathcal{J}&=&\left|  \frac{\partial \vec{x}_s}{\partial\theta}\times \frac{\partial \vec{x}_s}{\partial\varphi}  \right|; \\
 J_h &=& \frac{\partial \tilde{\mathcal{K}}_\varphi}{\partial\theta} - \frac{\partial \tilde{\mathcal{K}}_\theta}{\partial\varphi} 
\end{eqnarray}
to ensure that $\vec{\nabla}\cdot\vec{j}=0$.  The Heaviside step function $H(a-r)$ has minus the Dirac delta function $\delta(r-a)$ as its derivative, and $\mathcal{J}$ is the Jacobian of $(r,\theta,\varphi)$ coordinates. The net current flowing from the plasma to the wall $\oint \vec{j}\cdot\hat{n}\mathcal{J}d\theta d\varphi$ must be zero.  Since $\vec{\nabla}\theta\times\vec{\nabla}\varphi = \hat{n}/\mathcal{J}$, this implies $\oint J_hd\theta d\varphi =0$, which requires $\tilde{\mathcal{K}}_\theta$ and $\tilde{\mathcal{K}}_\varphi$ be periodic functions.  

The choice of the functions $\kappa$, $\tilde{\mathcal{K}}_\theta$, and $\tilde{\mathcal{K}}_\varphi$ can be made unique by choosing the single-valued part of the current potential, $\tilde{\kappa}$, so the normal magnetic field is all produced by $\tilde{\kappa}$, which separates the part of the magnetic field that can be produced by magnetic dipoles from the parts that cannot.  With this choice, the effects of $G_s,I_s,\tilde{\mathcal{K}}_\theta$, and $\tilde{\mathcal{K}}_\varphi$ on the magnetic field are limited.  $G_s$ affects only the toroidal magnetic flux enclosed by the shell.  $I_s$ affects only the poloidal magnetic flux outside of shell, which is the magnetic flux through the central hole of a torus.  

The current density $\tilde{\mathcal{K}}_\theta  \delta(r-a)(\hat{n}\times \vec{\nabla}\theta)$ is the toroidally flowing halo current in the shell.  The current density $\tilde{\mathcal{K}}_\varphi\delta(r-a)(\vec{\nabla}\varphi\times\hat{n})$ is the poloidally flowing halo current.   

The $\vec{j}\times\vec{B}$ force due to the halo current is can be determined using $\vec{\nabla}\varphi\times\hat{n} =(\partial\vec{x}/\partial\theta)/\mathcal{J}$ and $\hat{n}\times\vec{\nabla}\theta =(\partial\vec{x}/\partial\varphi)/\mathcal{J}$.   These three relations follow from the definition of the coordinate system, Equation (\ref{coord-system}), and the theory of general coordinates.



\section{Single-valued current potential $\tilde{\kappa}$ \label{sec:svcp}}

\subsection{Overview of the properties of $\tilde{\kappa}$}

It may be helpful to give the basic properties of the single-valued part of the current potential $\tilde{\kappa}(\theta,\varphi)$ that will be derived in this section.

 The force on the shell produced by $\tilde{\kappa}$ is given by Equation (\ref{shell force}).
  
The reduction in the normal magnetic field on a shell produced by the single-valued part of the current potential $\tilde{\kappa}$ will be shown to be given, Equation (\ref{kappa-S}), by the shielding coefficient $S_\kappa$, or more precisely by $1/S_\kappa$.  The shielding coefficient is defined so $1+S_\kappa$ is the externally driven normal field on the shell divided the actual normal field.  When the shielding is perfect,  $S_\kappa\rightarrow\infty$; the normal field on the shell is zero.  When there is no shielding  $S_\kappa=0$; the normal field on the shell equals the externally driven normal field.
 
 The single-valued part of the current potential is linearly dependent on the normal component of the magnetic field $B_\bot(\theta,\varphi) \equiv \hat{n}\cdot\vec{B}$ on the shell but has different forms depending upon whether electrically isolated tiles or a continuous conductor is being represented:

\begin{itemize}

\item The current potential due to electrically-isolated perfectly-conducting square tiles of width $w_t$ and depth $d_t$, which have a separation $\delta_t$, is derived in Section \ref{sec:tcp} and will be shown to be
 \begin{eqnarray}
&& \tilde{\kappa} = - s_t \frac{ B_\bot}{\mu_0},  \hspace{0.2in} \mbox{   where   } \label{tile cp}\\
&& s_t \equiv d_t \frac{ w_t}{\delta_t}. \label{tile s}
\end{eqnarray}
The distance $s_t$ is a geometrical property of the tiled surface.  When the tiles are $w_\ell\times w_s$ rectangles, $w_t$ in Equation (\ref{tile s}) is replaced by  $w_t=(w_\ell + w_s)/2$.  

Equation (\ref{kappa-S}) implies the shielding produced by the tiles is $S_t=ks_t/2$.

The resistivity of the tiles in ITER quickly broadens the effective separation of the tiles $\delta_e$ beyond their actual separation $\delta_t$.  When the normal field $B_\bot$, grows on a time scale $\tau_g$, the length $s_t$ will be shown to be
\begin{eqnarray}
 s_t &=& d_t\sqrt{\frac{\tau_t}{\tau_g}}\\
 \tau_t &\equiv& \frac{4\mu_0w_t^2}{\eta}
 \end{eqnarray}
 
 Magnetic probe measurements using the actual first wall modules could determine $s_t(\tau_g)$ empirically.  The derivation of the form of $s_t(\tau_g)$ given in Section \ref{sec:tcp} is useful as an estimate but is not a replacement for an actual measurement.
  
  \item The current potential of a continuous shell with resistivity $\eta$ and thickness $d_s$, which is penetrated by a normal field $B_\bot$ that grows on a time scale $\tau_g$ is derived in Section \ref{sec:ccp} and will be shown to be
  \begin{eqnarray}
\nabla^2\tilde{\kappa} = \frac{d_s}{\eta\tau_g}B_\bot . \label{cp-c}
\end{eqnarray}
  
  \end{itemize}
  
  A nearby conducting wall greatly reduces the force on either a tiled or a continuous shell.  This reduction is derived in Section \ref{sec:ncw}.


\subsection{Shielding given by $\tilde{\kappa}$  \label{sec:shielding}}

The shielding coefficient $S_\kappa$ produced by $\tilde{\kappa}$, the single-valued part of the current potential, can be calculated in a simple model in which the magnetic field that drives the current potential is given by $\vec{B}_d = \vec{\nabla}\phi_d$ with $\phi_d \propto e^{-kx} \cos(ky)$.  The shell is located on the $x=0$ plane.

This section will derive the relation between $\tilde{\kappa}$ and the shielding coefficient $S_\kappa$:
\begin{eqnarray}
\tilde{\kappa} =  - 2\frac{B_\bot}{\mu_0 k} S_\kappa. \label{kappa-S}
\end{eqnarray}
This section will also show that when a perfect conductor  is located a distance $\Delta$ behind the shell with $k\Delta<<1$, the current in the shell $\mathcal{I}$, which is given by Equation (\ref{I-kappa}), and hence the force on the shell are greatly reduced.

For generality, first consider the case of driving magnetic fields from both sides of the conducting shell, a normal field $B_n$ from the $x<0$ side and a normal field $B_p$ from the $x>0$ side.  
\begin{eqnarray}
\phi_n &=& - \frac{B_n}{k}\left( e^{-kx} + e^{kx}\right) + \frac{B_\bot}{k}e^{kx} \mbox{   for  } x<0 \hspace{0.3in}\\
&& \rightarrow - \frac{B_n}{k} e^{-kx} \mbox{   as  } x\rightarrow-\infty; \\
\phi_p &=&  \frac{B_p}{k}\left( e^{-kx} + e^{kx}\right) - \frac{B_\bot}{k}e^{-kx}  \mbox{   for  } x>0 \\
&& \rightarrow\frac{B_p}{k} e^{kx} \mbox{   as  } x\rightarrow+\infty; \\
B_\bot &=& \left(\frac{\partial \phi_{p/n}}{\partial x}\right)_{x=0}\cos(ky),
\end{eqnarray}
where $\phi_{p/n}$ means either $\phi_{p}$ or $\phi_{n}$, and $B_n$ and $B_p$ are proportional to $\cos(ky)$.  The jump in the magnetic potential across the shell is given by Equation (\ref{phi-jump}),
\begin{eqnarray}
(\phi_p-\phi_n)_{x=0} &=& \frac{2}{k}\left(B_n+B_p -B_\bot \right) \\
&= & -\mu_0 \tilde{\kappa},  \mbox{  so  } \\
S_\kappa &\equiv& \frac{B_n+B_p}{B_\bot} -1,
\end{eqnarray}
gives Equation (\ref{kappa-S}), which relates $\tilde{\kappa}$ to $S_\kappa$.  When there is no magnetic field driven from the $x>0$ side, $S_\kappa = (B_n/B_\bot) -1$, which gives the extent to which the normal component of the magnetic field is reduced by the shielding effect of the current potential

An important special case of a magnetic field being driven from the $x>0$ side is when there is a perfect conductor at $x=\Delta$, so $(\partial \phi_p/\partial x)_{x=\Delta}=0$.  The potential in the $x>0$ region that has the normal field $B_\bot$ at $x=0$ and zero at $x=\Delta$ is
\begin{eqnarray}
\phi_p &=& - \frac{B_\bot}{k}\frac{ e^{-kx}+e^{-2k\Delta}e^{kx} }{1-e^{-2k\Delta}},  \mbox{   so   } \\
S_\kappa &=& \frac{B_n}{B_\bot} - \frac{1}{1-e^{-2k\Delta}}.
\end{eqnarray}

The overall shielding, $S$, is the shielding due to the current potential and the shielding due to the perfect conductor at $x=\Delta$,
\begin{eqnarray}
S &=& S_\kappa + S_c \mbox{  where   }  \label{S=sum} \\
S_c &\equiv& \frac{e^{-2k\Delta}}{1-e^{-2k\Delta}} \\
&\rightarrow  & \frac{1}{2k\Delta} \mbox{   for } k\Delta\rightarrow0. \label{k Delta lim}
\end{eqnarray}

The reflection coefficient $\mathcal{R}$ gives the extent to which the magnetic field is reflected at the location of the shell by the effects of the shell and the perfect conductor behind it.  $\mathcal{R}$ is defined and is related to the overall shielding coefficient $S$ of the magnetic field by
\begin{eqnarray}
1-\mathcal{R} &\equiv& \frac{B_\bot}{B_n} \mbox{   and } \\
\mathcal{R} &=& \frac{S}{1+S}.
\end{eqnarray}

The nearby conductor reduces the current in the shell, $\mathcal{I}=-\partial\kappa/\partial y$, Equation (\ref{I-kappa}), and thereby the force. Using Equation (\ref{kappa-S}) for $\tilde{\kappa}$,
 \begin{eqnarray}
\mathcal{I} &=& - \frac{2S_\kappa}{k\mu_0} \frac{\partial B_\bot}{\partial y}, \mbox{   where  }\\
B_\bot &=& (1-\mathcal{R})B_n.   \mbox{   Let   }\\
 B_n&=& \bar{B}_n \cos(ky), \mbox{   then  } \\
\mathcal{I} &=& \frac{2S_\kappa}{1+ S_\kappa+S_c }  \frac{\bar{B}_d}{\mu_0}\sin(ky). \label{I-nearby-cond}
 \end{eqnarray}


\subsection{Current potential for a tiled shell \label{sec:tcp} }

The two jump conditions, Equations (\ref{B-perp jump}) and (\ref{B-tan jump}), together with the relation $\tilde{\kappa}=-s_tB_{\bot}/\mu_0$ give the magnetic field components on the two sides of a tiled shell when a magnetic field of amplitude $B_n$ is driven from the negative $x$ side of a planar shell.
\begin{eqnarray}
&&\mbox{The field on the  } x<0 \mbox{    side is} \nonumber\\
&& \frac{B_x}{B_n} =\frac{  \cos(ky)}{1+\frac{1}{2}ks_t } , \hspace{0.1in}  \frac{B_y}{B_n} =\frac{1+ks_t}{1+\frac{1}{2}ks_t} \sin(ky). \nonumber\\
\nonumber\\
&&\mbox{The field on the } x>0 \mbox{    side is}  \nonumber\\
&&  \frac{B_x}{B_n} =\frac{ \cos(ky)}{1+\frac{1}{2}ks_t},  \hspace{0.2in}  \frac{B_y}{B_n} =\frac{ \sin(ky)}{1+\frac{1}{2}ks_t}. \label{tiled field}
\end{eqnarray}
The current potential of a tiled shell is equivalent to magnetic dipoles oriented perpendicular to the shell.  In a magnetized shell, the dipoles are oriented tangentially and the magnetic field on the two sides has a different form, Equation (\ref{B-bot jump}).

\subsubsection{Derivation of the tiled-shell $\tilde{\kappa}$}

This section will assume the tiles are electrically isolated and perfect conductors.  The effect of resistivity will be determined in Section \ref{sec:tile resis}.  

The solution for the magnetic field in the presence of perfect conductors is a minimum of the magnetic energy.  The heuristic argument for this principle is that if a system consists of perfect, $\eta=0$, conductors embedded in resistive material, $\eta\neq0$, then the currents in the resistive material will dissipate magnetic energy until the magnetic field reaches a curl-free state outside the perfect conductors.  If a unique state exists it must be a state of minimum magnetic energy.  There are formal proofs \cite{Perfect conductors:2013}, which include the effect of ideal skin currents of width $c/\omega_{pe}$ on the perfect conductors.

Consider a sinusoidal magnetic field $\vec{B}_d$ that is driven from a great distance $X$ in the negative $x$ direction, with a thin tiled shell located at $x=0$. The magnetic scalar potential and the magnetic field for $0>x>-X$ are
 \begin{eqnarray}
 \phi &=&- \frac{B_n}{k}\cos(ky)\Big(e^{-kx} +\mathcal{R} e^{kx}\Big) \label{driven-potential} \\
 \vec{B} &=&B_n\Big\{\cos(ky)\Big(e^{-kx} -\mathcal{R} e^{kx}\Big)\hat{x}   \nonumber\\
 &&+\sin(ky)\Big(e^{-kx} +\mathcal{R} e^{kx}\Big)\hat{y}\Big\}. \label{B-neg}
  \end{eqnarray}
  The constant $\mathcal{R}$ is the extent to which the tiled shell reflects the magnetic field.   The energy averaged over $ky$ in the negative $x$ region  over a width in $z$ equal to the width of a tile $w_t$ is
  \begin{equation}
  W_- = \frac{B_n^2}{2\mu_0} \frac{w_t^2}{2k} \left( e^{2kX} + \mathcal{R}^2 -1\right).
  \end{equation}

  The magnetic scalar potential and the magnetic field for $x>0$ are 
 \begin{eqnarray}
\phi &=& - (1-\mathcal{R})\frac{B_n}{k}\cos(ky)e^{-kx}\\
  \vec{B} &=&(1-\mathcal{R})B_n\Big\{\cos(ky) e^{-kx}\hat{x} \nonumber\\
 &&+\sin(ky)e^{-kx} \hat{y}\Big\}.  \label{B-pos}
 \end{eqnarray}
 
 Since $\Big[B_y\Big] =-2\mathcal{R} B_n\sin(ky)= -\mu_0\partial \tilde{\kappa}/\partial y$,
 \begin{equation}
\tilde{\kappa} = -\frac{2\mathcal{R}}{\mu_0} \frac{B_n}{k}\cos(ky). \label{kappa first}
 \end{equation}
 
 The magnetic flux going through the tiled shell per tile is $B_x(x=0+)w_t^2 = (1-\mathcal{R}) B_n \cos(ky)w_t^2$, but all of this flux must slip through the channels of width $\delta_t$, which occupy only a small fraction of the area of the shell, $2\delta_tw_t$.  Consequently, the magnetic field in the channels must be
 \begin{eqnarray}
 \vec{B}_{ch} &=& (1-\mathcal{R})B_n \cos(ky)\frac{w_t}{2\delta_t} \hat{x}, \hspace{0.2in}\mbox{and} \\
 W_{ch} &=& (1-\mathcal{R})^2\frac{B_n^2}{2\mu_0}d_t w_t^2 \frac{w_t}{4\delta_t} 
 \end{eqnarray}
 is the energy in the channel per tile after averaging over $\cos^2(ky)$.  The depth of the channels between the tiles is $d_t$.  It has been assumed $d_t>>w_t$, so the fringing fields that occur near the entrance and exit of the channels do not contribute significantly to the energy $W_{ch}$.
 
The energy in the channels between tiles can be written as
 \begin{eqnarray}
W_{ch} &=& (1-\mathcal{R})^2 \frac{B_n^2}{2\mu_0}\frac{w_t^2}{2k}  S_t \\
S_t&\equiv& \frac{ kd_t}{2} \left(\frac{w_t}{\delta_t}\right). \label{S-t def}
\end{eqnarray}
$S_t$ will be found to be the shielding coefficient of the tiled surface.

In the limit $S_t>>1$, the shielding becomes complete $\mathcal{R}\rightarrow1$, and the only magnetic energy outside of the shell is in the region $0>kx>-kX$.  
The actual reflection coefficient $\mathcal{R}$ is the one that minimizes the total magnetic energy
\begin{eqnarray}
W_-+W_{ch} &=&\frac{B_n^2}{2\mu_0} \frac{w_t^2}{2k} \Big( e^{2kX}+  \mathcal{R}^2 -1 \nonumber\\
&&+ (1-\mathcal{R})^2  S_t\Big), \mbox{    which is   }  \hspace{0.2in} \\
 \mathcal{R}&=&\frac{S_t}{1+S_t}.
\end{eqnarray}
Using Equation (\ref{kappa first}) and $B_\bot=(1-\mathcal{R})B_n$,
\begin{eqnarray}
\mu_0\tilde{\kappa}&=&  \frac{2\mathcal{R}}{1-\mathcal{R}}\frac{B_\bot}{k} \\
 &=&- 2\frac{S_t}{k} B_\bot. \label{kappa-exp}
 \end{eqnarray}
The definition of $S_t$, Equation (\ref{S-t def}) shows that Equation (\ref{kappa-exp}) is equivalent to Equations (\ref{tile cp}) and (\ref{tile s}) for the current potential of a tiled surface.

 
  \subsubsection{Effect of tile resistivity \label{sec:tile resis} }
 
 The tiles are not perfect conductors so the effective widths of the channels between the tiles grows due to tile resistivity. When the resistivity enhanced distance between tiles is larger than the actual distance, Equations (\ref{tile cp}) and (\ref{tile s}) are modified.
 \begin{itemize}
 \item When $B_\bot\propto e^{t/\tau_g}$, it will be shown that
 \begin{eqnarray}
 \tilde{\kappa} &=& - d_t \sqrt{\frac{\tau_t}{ \tau_g}} \frac{ B_\bot}{\mu_0} \hspace{0.2in} \mbox{where   }\\
  \tau_t &\equiv& \frac{\mu_0w_t^2}{4\eta}. \label{tau-t eq}
\end{eqnarray}

\item When $B_\bot = \sqrt{2}\bar{B}_\bot\cos(\omega t)$, where $\bar{B}_\bot$ is the averaged normal field, $\bar{B}_\bot^2 \equiv \Big<B_\bot^2\Big>$, it will be shown that
\begin{eqnarray}
\tilde{\kappa} &=& - d_t  \sqrt{\tau_t\omega} \frac{ \bar{B}_\bot}{\mu_0}\cos\left(\omega t-\frac{\pi}{4}\right)
\end{eqnarray}

\end{itemize}
 
 The widening of the effective channel width can be calculated using the $z$ component of the vector potential, which obeys the evolution equation $\nabla^2A=(\eta/\mu_0) \partial A/\partial t$.  If the magnetic perturbation is growing on a time scale $\tau_g$, then the vector potential has the form in the tiles $A\propto e^{-Ky} e^{t/\tau_g}$, so $K^2=(\mu_0/\eta)/\tau_g$.  The effective channel width is $\delta_e=\delta_t + 2/K$, so
 \begin{equation}
 \frac{1}{\delta_e}=\frac{K}{2 +K\delta_t}.
 \end{equation}  Since the actual channels are very narrow, the effective channel width $\delta_e=2/K$ and
 \begin{eqnarray}
 \delta_e &=& w_t \sqrt{\frac{4\eta\tau_g}{\mu_0w_t^2}} \\
 &=& w_t\sqrt{\frac{\tau_g}{\tau_t}};  \label{delta-e equation}\\
 s_t &=& d_t \frac{w_t}{\delta_e} \\
 &=& d_t\sqrt{\frac{\tau_t}{\tau_g}}.
 \end{eqnarray}
 
 The ITER first-wall tiles are beryllium which has a resistivity $\eta = 36\times10^{-9}~\Omega\cdot$m, so $\eta/\mu_0=2.9\times10^{-2}~$m$^2$/s.  They are backed by a copper alloy, which will be assumed to be electrically a part of the tile and to have a comparable resistivity.  Assuming $w_t\approx 5~$cm, the tile time constant, Equation (\ref{tau-t eq}), is $\tau_t\approx$ 20~ms. 
  
  For an oscillatory field, the vector potential is the real part of $A\propto e^{-Kx} e^{i\omega t}$ and $K=\sqrt{i} \sqrt{\omega\mu_0/\eta}$, where $\sqrt{i}=(1+i)/2$.  The normal component of the magnetic field is proportional to $\cos(\omega t)$.  The real part of $e^{i\omega t} Kw_t/2$ is $\sqrt{\omega\mu_0/16\eta}\Big(\cos(\omega t)-\sin(\omega t)\Big)=\sqrt{\omega\mu_0/8\eta}\cos(\omega t-\pi/4)$.
  
  
\subsection{Effect of a nearby conducting wall on the force on tiles \label{sec:ncw}}

The force on a surface covered by resistive tiles and  separated from a nearby perfect conductor by a distance $\Delta$  will be shown to be equivalent to that of a perfect conductor when the wavenumber $k$ of the normal field to the surface $B_\bot$ is larger than $k_{eq}$, where 
\begin{eqnarray}
\frac{1}{k_{eq}} &\equiv& \sqrt{d_t\Delta}\left(\frac{\tau_t}{\tau_g}\right)^{1/4},
\end{eqnarray}
$d_t$ is the depth of the tiles, $\tau_t\sim20~ms$ is given by Equation (\ref{tau-t eq}) and $\tau_g$ is the growth rate of $B_\bot$.  For for smaller $k$, the force drops as $(k/k_{eq})^2$.

To show this, note that the force on the tiles is given by surface current $\vec{\mathcal{I}}$ crossed with the magnetic field at the shell, Equation (\ref{mid-eq}).  $\vec{\mathcal{I}}$ due to the tiles is in the direction orthogonal to both the normal direction $\hat{n}$ and the direction $y$ in which $B_\bot$ varies.  When a conductor is nearby, $k\Delta<<1$, the shielding effect of the conductor is given by Equation (\ref{k Delta lim}), $S_c=1/(2k\Delta)$.  Equation (\ref{I-nearby-cond}) implies that when $S_c>>S_t$,
\begin{eqnarray}
\mathcal{I} &=& \frac{ S_t}{S_c}\frac{2\bar{B}_n}{\mu_0}\sin(ky).
\end{eqnarray}
$S_t$ is given by Equation (\ref{tile s}) with the actual tile separation $\delta_t$ replaced by the effective separation $\delta_e$, Equation (\ref{delta-e equation}), so
\begin{eqnarray}
\mathcal{I}&=& k^2 \Delta \frac{d_tw_t}{\delta_e}\frac{2\bar{B}_n}{\mu_0}\sin(ky) \\
&=& \left( \frac{k}{k_{eq}}\right)^2 \frac{2\bar{B}_n}{\mu_0}\sin(ky),.
\end{eqnarray}
For wavenumbers larger than $k_{eq}$, Equation (\ref{I-nearby-cond}) implies the first wall is acted upon by essentially the same force as if it were a perfect conductor, for then $S_t >S_c>>1$.  For smaller wave numbers the force drops as $(k/k_{eq})^2$.
  
When the tile time constant is $\tau_t\approx$ 20~ms, the separation from a perfect conductor is $\Delta\approx 0.5$~m and the thickness of the tiled structures is $d_t\approx 2.5$~cm,
  \begin{eqnarray}
 \frac{1}{k_{eq}}&\approx& \frac{0.24}{\tau_g^{1/4}}  \label{k_eq}
  \end{eqnarray}
  in units of meters for $1/k_{eq}$ and milliseconds for $\tau_g$.
  


\subsection{Current potential for a conducting shell \label{sec:ccp}}

The current potential for a conducting shell is calculated by assuming the external driving magnetic field has only $x$ and $y$ dependence and has the same dependencies as given in Equations (\ref{driven-potential}) through (\ref{B-pos}).  The current density within the shell is $\vec{j}=j_z\hat{z}$, and $j_z$ has no $x$ dependence across the thin shell of thickness $d_s$.  

 Ampere's law, $\partial B_y/\partial x =\mu_0 j_z$, implies the jump in $B_y$ across the shell is $[ B_y ] = \mu_0 j_z d_s$, which can be written using  Equations (\ref{B-neg}) and (\ref{B-pos}) as $\mu_0 j_z d_s = 2\mathcal{R} B_n \cos(ky)$.  

Faraday's and Ohm's law imply $\partial \vec{B}/\partial t = - \vec{\nabla}\times( \eta j_z\hat{z})$ so $B_x/\tau_g= k \eta j_z$.  

The $x$-component of the field at $x=0$ is $B_x= (1-\mathcal{R}) B_n \cos(ky)$.  The relation between the reflection coefficient $\mathcal{R}$ and the shielding coefficient for a continuous shell is $R/(1-R) =S_{cs}$, so
\begin{eqnarray}
S_{cs} &=&  \frac{\mu_0 d_s}{2\eta\tau_g k}. \label{S-cs}
\end{eqnarray}
Equation (\ref{kappa-S}), which relates $\tilde{\kappa}$ to the shielding coefficient, implies Equation (\ref{S-cs}) is equivalent to Equation (\ref{cp-c}) for the current potential of a shell that is a continuous conductor.

  

\section{Non-single-valued current potentials \label{sec:nsvcp} }

\subsection{Toroidal flux relaxation \label{sec:tfr} }

The relation between the enclosed toroidal magnetic flux $\psi_t$ and the net poloidal current $G_s$ in a shell can be studied by considering a thin cylindrical shell which is periodic in $z$ by letting $\varphi=z/R_0$ be the ``toroidal" angle with $2\pi R_0$ the period.   

The shell has a radius $r_s$ and thickness $d_s$, which encloses an area $A_s=\pi r_s^2$.  The $z$-directed magnetic field has a toroidal flux $\psi_t=B_{in}(t) A_s$ with $B_{in}$ the spatially-constant toroidal field within the shell.  The poloidal current density in the shell $j_\theta = G_s/(2\pi R_0 d_s)$.

Faraday's law coupled with Ohm's law implies
\begin{eqnarray}
\frac{\partial B_z}{\partial t} &=& - \frac{1}{r}\frac{\partial}{\partial r}\left(r\eta j_\theta\right), \mbox{   so   } \label{d toroidal/dt}\\
\frac{\partial \psi_t}{\partial t}&=& - \mathcal{R}_p G_s ,  \mbox{     where   } \label{dot(psi-t)} \\
\mathcal{R}_p &\equiv& \frac{\eta 2\pi r_s}{2\pi R_0 d_s} 
\end{eqnarray}
is the poloidal resistance of the shell.

Ampere's law implies
\begin{eqnarray}
\frac{\partial B_z}{\partial r} &=& -\mu_0 j_\theta \mbox{  so  } \\
\left[ B \right] &\equiv& B_{out}-B_{in} \\
&=& - \frac{ \mu_0 G_s }{2\pi R_0}, \mbox{    where    } \label{B_z Ampere}
\end{eqnarray}
$B_{out}$ is the toroidal field outside the shell.

To study the resistive relaxation of the toroidal flux, assume the is a perfectly conducting outer cylinder that encloses an area $A=\pi r_o^2$ with $r_o>r_s$.  The toroidal flux $\Psi_t$ enclosed by $A$ is conserved.  The average magnetic field inside the larger cylinder $\bar{B}\equiv\Psi_t/A$ is independent of time, and 
\begin{eqnarray}
 \left[ B \right] &=& (\bar{B} - B_{in})\frac{A}{A-A_s},  \mbox{    or   } \\
 B_{in}&=& \bar{B}+ \frac{A-A_s}{A} \frac{\mu_0G_s}{2\pi R_0},   \mbox{   and  }\\
 \psi_t &=& \frac{A_s}{A}\Psi_t + \frac{(A-A_s)A_s}{A} \frac{\mu_0G_s}{2\pi R_0}\\
  &=&  \frac{A_s}{A}\Psi_t + L_p G_s \mbox{  with   } \label{psi-t - G_s} \\
  L_p&=& \mu_0 \frac{(A-A_s)}{A} \frac{A_s}{2\pi R_0} \\
  &=& \frac{\mu_0 r_s}{2}  \left(1-\frac{r_s^2}{r_o^2} \right)\frac{r_s}{R_0}
 \end{eqnarray}
 an inductance coefficient.  Equation (\ref{dot(psi-t)}) for the time derivative of the toroidal flux $\psi_t$ inclosed by the shell and Equation (\ref{psi-t - G_s}), which relates $\psi_t$ and $G_s$ imply
 \begin{equation}
\frac{\partial G_s}{\partial t} = - \frac{G_s}{L_p/\mathcal{R}_p}.
\end{equation}

When the magnetic field and the wall have exact toroidal symmetry, there is no normal magnetic field to the wall, and the result given above can be generalized for a shell of arbitrary shape.  Equation(\ref{B_z Ampere}) generalizes trivially.  Equation (\ref{d toroidal/dt}) is generalized using the vector-calculus identity that switches the area integral of a curl into a line integral $\int (\vec{\nabla}\times \vec{E})\cdot d\vec{a} =\oint \vec{E}\cdot d\vec{\ell}$.  The poloidal loop voltage is $V_p\equiv\oint \vec{E}\cdot d\vec{\ell}$.  The implication is that the change in the toroidal flux, $\psi_t =\int \vec{B} \cdot d\vec{a}$ enclosed by a curve is the loop voltage around the curve.  The poloidal loop voltage is the poloidal resistance $\mathcal{R}_p$ times the poloidal current $G_s$.  

There are two loop voltages on a topologically toroidal surface.  The time derivative of the toroidal flux enclosed by a curve is $d\psi_t/dt=-V_p$, where the poloidal loop voltage is for a poloidal circuit of the torus.

\vspace{0.2in}


\subsection{Poloidal flux relaxation \label{sec:pfr} }

Poloidal flux relaxation can also be studied in a cylinder, but poloidal and toroidal fluxes have a fundamental difference.  The poloidal magnetic field depends on currents at radii smaller than the radius one is studying, and the poloidal flux is determined by the poloidal field at larger radii.  For the toroidal magnetic field and flux the two statements about radii are the other way around.  

The toroidal net current in the shell $I_s$ controls the entry of external poloidal flux $\Psi_p$ into the region enclosed by the shell.  The toroidal current density in the shell is $j_z = I_s/(2\pi r_s d_s)$ using the same notation as in Section \ref{sec:tfr} to characterize the cylindrical geometry.  

The poloidal flux between the radius of the shell $r_s$ and the outer conducting wall $r_o$ is given by the curl-free poloidal field in that region $B_\theta=(r_s/r)B_\theta(r_s+)$, where $B_\theta(r_s+)$ is the poloidal field just outside the shell.
\begin{eqnarray}
\Psi_p &=& 2\pi R_0 \int_{r_s}^{r_o} \frac{r_s}{r} B_\theta(r_s+) dr\\
&=& 2\pi R_0 r_s \ln\left(\frac{r_o}{r_s}\right) B_\theta(r_s+).
\end{eqnarray}

Assuming there are no toroidal currents within the region enclosed by the shell, $B_\theta=0$ for $r<r_s$, and Ampere's law implies
\begin{eqnarray}
\frac{1}{r}\frac{\partial rB_\theta}{\partial r} &=& \mu_0 j_z \mbox{  so  } \\
j_z &=& \frac{B_\theta(r_s+) }{\mu_0 d_s},  \\
B_\theta(r_s+) &=& \frac{\mu_0 I_s}{2\pi r_s},  \mbox{   and   }\\ 
\Psi_p &=& L_t I_s \mbox{   where   }\\
L_t &=& \mu_0 R_0 \ln\left(\frac{r_o}{r_s}\right)
\end{eqnarray}
is the toroidal inductance

The combination of Faraday's law and Ohm's law gives
\begin{eqnarray}
\frac{\partial B_\theta}{\partial t}&=&-\frac{\partial \eta j_z}{\partial r}  \mbox{  so  } \\
j_z &=& -\frac{1}{\eta} \frac{\partial }{\partial t} \int_{r_s}^{r_o} B_\theta dr \\
&=&- \ln\left(\frac{r_o}{r_s}\right) \frac{ r_s}{\eta}\frac{\partial B_\theta(r_s+)}{\partial t}\\
&=& - \frac{1}{2\pi R_0\eta}\frac{\partial\Psi_p}{\partial t}; \\
\frac{\partial\Psi_p}{\partial t}&=& - \mathcal{R}_t I_s,  \mbox{   where  } \\
\mathcal{R}_t &\equiv& \frac{\eta 2\pi R_0 }{2\pi r_s d_s},  \mbox{   and  } \\
\frac{\partial I_s}{\partial t}&=&- \frac{I_s}{L_t/R_t},
\end{eqnarray}
where $\mathcal{R}_t$ is the toroidal resistivity.  The toroidal loop voltage in the shell is $V_t =-\partial\Psi_p/\partial t$.   

The toroidal resistivity is an aspect ratio squared, $(R_0/r_s)^2$, times larger than the poloidal resistivity, but there is also a factor of the aspect ratio squared between the two inductances.  The ratio of the two $L/\mathcal{R}$ times, 
\begin{equation}
\frac{\tau_p}{\tau_t} = \frac{2\ln\left(\frac{r_o}{r_s}\right)}{1-\frac{r_s^2}{r_o^2}},
\end{equation}
depends only on the ratio $r_s/r_0$ of the radii of the two cylindrical surfaces, the shell and the outer perfect conductor.


\section{Implications for the study of off-normal events \label{sec:off normal} }

Concerns about major machine damage in tokamaks focus on the effects of off-normal events: disruptions, vertical displacements, resistive wall modes, and runaway electrons.  These effects are reviewed in the ITER Physics Basis Documents \cite{ITER physics,Hender:2007}, but more recent developments also need to be considered.  

We consider effects due to magnetic forces, which could be greatly enhanced by electrical shorts between the beryllium tiles covering the first wall.  A possible cause of shorting is tile melting \cite{melting:2002,melting:2014}.  The actual melt damage seen on the JET ITER-like Wall and divertor is illustrated and discussed in \cite{JET-melt}.

The focus of this paper is not on tile shorting but on the magnetic behavior of the chamber walls on short time scales $\lesssim50~$ms.  This is faster than the presently envisioned time for the quench in the plasma current, but faster current quenches might be beneficial for controlling runaway electrons.  In any case, the interaction of the walls with rapidly changing magnetic perturbations is important during rapid drops in the plasma pressure, called a thermal quench, and due to vertical displacement events and resistive wall modes driving both halo and induced currents.

The shorter the time scales, the more important are the dipolar currents associated with the beryllium tiles since they are the closest conducting surface to the plasma.  Dipolar currents arise to impede any change in $B_\bot$, the magnetic field component  that is normal to the tiled surface.  Since the tile structures are thin, the effect can be modeled by two jump conditions:  $\Big[ B_\bot \Big] =0$ and $\Big[ \vec{B}_{tan} \Big] =\vec{\nabla}(s_tB_\bot)$, where $\vec{B}_{tan}$ is the part of the magnetic field that is tangential to the tiled surface.  The coefficient that measures the strength of the effect has units of meters and is given by $s_t=d_tw_t/\delta_e$, where $d_t$ is the thickness of the tiles, $w_t$ the width of square tiles, and $\delta_e$ the effective separation between tiles.  The effective separation is determined by resistive penetration; $w_t/\delta_e = \sqrt{\tau_t/\tau_g}$, where the resistive time scale of the tiles is $\tau_t \equiv \mu_0 w_t^2/4\eta\approx 20~$ms.  The normal field is assumed to grow as $B_\bot\propto e^{t/\tau_g}$.  The force that is exerted on the tiled surface is given by Equation (\ref{shell force}) with the current potential  $\kappa$ given by $-s_t B_\bot/\mu_0$.  The force is particularly strong when the current potential has a strong toroidal variation.

The drop in the plasma pressure, $P_{pl}$ is generally expected to require approximately a millisecond in ITER, but the drop in the electron pressure could be much faster.  In a multi-kilovolt plasma, the electron temperature should drop very quickly if the magnetic surfaces break \cite{Boozer:j flattening}, because the mean-free-path of the electrons can be as long as the connection length of chaotic field lines to the outside world, about a hundred toroidal transits.  The rate of loss of the electron thermal energy would be given by the time it takes electrons moving along the chaotic magnetic field lines to cross the plasma, $\sim 50~\mu$s as seen on DIII-D \cite{Paz-Soldan:2020}, until the electron temperature drops to approximately 1~keV.  At lower temperatures, collisions impede the energy transport.  It will take much longer for the ions to equilibrate, but the plasma pressure should quickly drop by a factor of two.  If it does not, it tells us something about the state of magnetic chaos during the thermal quench.  The hundred transits is comparable to the independent observations in a numerical simulation of a tokamak disruption by Valerie Izzo \cite{Izzo:2020} and that by Eric Nardon et al., which is not yet published.

The forces associated with extremely fast time scales are exerted on the nearest surface that can carry the current, which is the tiled surface for currents associated with changes in $B_\bot$ but is more complicated for currents that change the tangential field $\vec{B}_{tan}$ but not $B_\bot$.   

The overall effect of the force on a blanket module that is associated with extremely rapid magnetic change is reduced by inertia.  Fast-acting forces can be treated as impulses for time scales shorter than the inertial time, which is of order milliseconds.  An inertial time is required for a disruption force to displace an ITER blanket module by the maximum tolerable distance  \cite{ITER:blanket2014}, $\xi=3$~mm.  The inertial time  is $\sqrt{2 (M_b/P_dA_b)\xi}$, where $M_b=4.6~$tons is the mass and $A_b=1.5~$m$^2$ is the frontal area of a module, and $P_d$ is the disruption force per unit area.  Expressing $P_d$ in atmospheres, the inertial time is $(4/\sqrt{P_d})$~ms.  A typical inertial time is a few milliseconds since a typical $P_d$ will be shown to be of order an atmosphere.

Any drop in the plasma pressure will drive both a poloidal and a toroidal current in the surrounding chamber.  The poloidal current can be calculated using the average equilibrium equation, Equation (D19) of Kruskal and Kulsrud \cite{K-K}.  This equation implies that if  the plasma filled the toroidal chamber then the change in the poloidal current $\delta G=2\pi R  \Big< \delta P_{pl}\Big>/B_\varphi$, where $R$ is the major radius and $ \Big<\delta P_{pl}\Big>$ is the volume-averaged change in the plasma pressure.  This change in poloidal current interacting with the toroidal field would produce a radial force per unit area on a conducting structure of $\Big<\delta P_{pl}\Big>$.  The pre-thermal-quench pressure of a hydrogenic plasma in atmospheres is $P_{pl} =3.17 n_eT$, where the electron density $n_e$ is in units of $10^{20}$m$^{-3}$ and the temperature of each species $T$ is in units of $10^4$~eV, which are the approximate density and temperature expected in ITER.   A drop in the  plasma pressure implies an inward force.  Without extensive shorting, a net poloidal current cannot flow in the first wall.  Nevertheless, a force can be exerted on the tiled first-wall surface particularly where toroidal symmetry is broken, Section \ref{sec:tfr}.  

In addition to forces, a fast drop in the electron pressure will produce a very large poloidal loop voltage, Section \ref{sec:tfr}.  The change in the toroidal flux is $\delta\psi_t=2\delta\beta B_\varphi A_{pl}$, where $A_{pl}\simeq 20~$m$^2$ is the cross-sectional area of the ITER plasma and $\beta_ = 2\mu_0 \Big<P_{pl}\Big>/B^2$.  Approximately two Webers of flux change would be produced by $\Big<\delta\beta\Big>=1\%$.  When this occurs over 50~$\mu$s, the poloidal loop voltage would be 40,000~V.

A toroidal current with a $\cos\theta$ distribution also arises in the chamber on the time scale of the thermal pinch due to the change in the required vertical field.  This current is smaller than the plasma current by approximately the inverse aspect ratio times the change in the poloidal beta, $\beta_\theta = 2\mu_0 \Big<P_{pl}\Big>/B_\theta^2. $  The change in the vertical field produces a normal magnetic field not only because the vertical field has a normal component but also because of poloidal and toroidal asymmetries.  Toroidal asymmetries that have a short spatial scale produce an especially strong local force on the tiled shell, Equation (\ref{shell force}), because the shell currents then interact with the toroidal magnetic field.  A careful study is required to determine whether there are any locations on the first wall at which a force of a dangerous amplitude arises.  

A resistive wall mode arises if the plasma would be ideally unstable, usually to a perturbation with a toroidal mode number of unity, without a nearby conducting structure but stable in the presence of that structure \cite{RWM:Boozer}.  In this situation, the non-axisymmetry of the equilibrium grows on a time scale determined by the resistance of the current path through the conducting structure.  The interaction with the tiled surface is complicated, especially with plasma rotation, and the growth rate can be high.  Careful studies are required to determine whether a significant danger can be posed to the first wall.  The large separation between the first wall and the conducting vessel, $\Delta\approx 0.5$~m, makes careful calculations particularly important.

The vertical displacement that normally follows a disruption has traditionally been assumed to evolve on the time scale of the surrounding conducting wall, which is slow, but it actually evolves on the current quench time when the wall is a perfect conductor \cite{Breizman:2018} and can even become unstable when the pre-disruption axisymmetric shaping is too strong \cite{Ideal VDE,Clauser:2020}.  The interaction with a tiled shell rather than a wall with continuous conductivity has not been studied.

The vertical displacement instability and the resistive wall mode become more threatening when the outer layers of the plasma are lost faster than the current decays, so the safety factor at the plasma edge reaches $q=2$ or $q=1$.  Plasmas with edge safety factors of $q=2$ or $q=1$ are so ideally unstable that a current must flow along the magnetic field lines in the plasma halo from one contact point of the plasma with the wall to another  \cite{Halo:Boozer,Halo}.  This halo current closes its path by flowing through the walls, Section \ref{sec:halo}.  The interaction, with or without halo current rotation, with a tiled wall has not been studied.  The study is made more subtle because the plasma intercepting the chamber walls may push plasma into the spaces between the tiles and produce shorting. 

Disruptions and runaways are actually three coupled issues---a solution to one tends to complicate the solution to the other two: (1) loss of position control of the plasma, (2) excessive power deposition, particularly on the divertor, and (3) wall melting due to runaway electrons.  All three issues must be carefully studied obtain an adequate compromise among them.  The magnetic interaction of the closest conduction structure to the plasma, the surface of beryllium tiles, needs to be understood to assess the effects of the loss of position control.  The faster the time scales and the shorter the wavelengths of the magnetic perturbations, the more important the tiled surface becomes.  The large number of individual tiles, $\sim26,000$, makes direct magnetics calculations intractable, but also makes the current-potential model of a tiled surface a credible approximation.

\vspace{0.2in}

\section*{Acknowledgements}

This material is based upon work supported by the U.S. Department of Energy, Office of Science, Office of Fusion Energy Sciences under Award Numbers DE-FG02-03ER54696, DE-SC0018424, and DE-SC0019479.   

\section*{Data availability statement}

Data sharing is not applicable to this article as no new data were created or analyzed in this study.


\appendix


\section{Current-potential approximation \label{sec:approx}}

As discussed in Section \ref{sec:cp}, the current potential $\kappa$ in a thin shell provides expressions for (1) the jump in the magnetic field normal to the shell from one side to the other, $\big[\vec{B}\cdot\hat{n}\big]$, (2) The jump in the tangential magnetic field, $\vec{B}_{tan} \equiv \vec{B}- \hat{n}\vec{B}\cdot\hat{n}$, from one side to the other, $\big[\vec{B}_{tan}\big]$, and (3) the integral across the shell of the current density tangential to the shell, $\vec{\mathcal{I}}$.  

This section derives the two conditions on the thickness of the shell, $d_s$, for the for current-potential approximation to be valid.  The first is a geometric constraint that the typical wavenumber of the magnetic field external to the shell, $k_t\sim \big| \vec{\nabla}\vec{B}\big|/B$, satisfies $k_td_s<<1$.  The second is a temporal constraint on the rate change $1/\tau_g$ of magnetic perturbations $\delta\vec{B}\propto e^{t/\tau_g}$.  The time $\tau_g$ must be long compared to the resistive time of the shell,
\begin{equation}
\tau_s \equiv \frac{\mu_0 d_s^2}{\eta}
\end{equation}
for the approximation $\big[\vec{B}\cdot\hat{n}\big]=0$ to be accurate.

The geometric constraint is obvious---the shell cannot be viewed as thin unless it is thin compared to the spatial scale over which the magnetic field varies---but the temporal constraint $\tau_g>>\tau_s$ is more subtle.  

A thin shell strongly shields a magnetic field as long as the radial wavenumber of the perturbation within the shell $K=\sqrt{\tau_s/\tau_g}/d_s$, Equation (\ref{K-def}), satisfies $K>>k$---even when $K d_s<<1$.  The thin shell approximation is accurate and useful when the thickness of the shell satisfies $d_s << 1/K<<1/k$.  This section uses a simple model to explore the properties of shells when $d_s<<1/k$ when $K>>k$---without an assumption on the size $K d_d$.

The evolving part of a magnetic field can be studied using the vector potential, $\vec{A}=A_z\hat{z}$, current density $\vec{j}=j_z\hat{z}$, and electric field, $\vec{E}=E_z\hat{z}$, which have the forms  
\begin{eqnarray}
A_z&=&A(x) \sin(ky) e^{t/\tau_g}, \\
\mu_0 j_z &=& - \nabla^2A_z, \\
&=& \mu_0 j(x) \sin(ky) e^{t/\tau_g}, \mbox{   and  } \\
E_z &=& -\frac{\partial A_z}{\partial t}=\eta j_z 
\end{eqnarray}
in $x,y,z$ Cartesian coordinates.  Within the conducting shell $|x|<d_s/2$, 
\begin{eqnarray}
A(x) &=& A_{-}e^{-Kx} - A_{+} e^{Kx}   \\
K^2 &=& k^2 + \frac{\mu_0}{\eta \tau_g} \nonumber\\
&\rightarrow& \frac{\mu_0}{\eta \tau_g} \label{K-def}
\end{eqnarray}
as $K/k\rightarrow\infty$.

The field is driven from large negative $x$, so in the insulating region $x<-d_s/2$, the vector potential is
\begin{eqnarray}
A(x) = \frac{\bar{B}_d}{k} \left( e^{-kx} - \mathcal{R} e^{kx}\right),
\end{eqnarray}
where $\bar{B}_d \sin(ky) \exp(t/\tau_g)$ would be the normal component, which means the $x$-component, of the field at the location of the shell if the shell were a perfect insulator.  $\mathcal{R}$ is the fraction of that field the shell reflects.  Since $kd_s<<1$, the magnetic field would change negligibly across the shell if it were a perfect insulator.  Behind the shell, $x>d_s/2$,  
\begin{eqnarray}
A(x) = A_b  e^{-kx}.
\end{eqnarray}

The two jump conditions on the boundaries of the conductor, $x=\pm d_s/2$ are $[A]=0$ and $[dA/dx]=0$.  In the limit $kd_s\rightarrow0$, these conditions can be given using
\begin{eqnarray}
&& E\equiv e^{Kd_s/2}:\\
&&A_{-} E  - A_{+} E^{-1}  = \frac{\bar{B}_d}{k} \left( 1 - \mathcal{R} \right), \label{c1} \\
&& A_{-} E +A_{+}  E^{-1}  = \frac{\bar{B}_d}{K}   \left( 1 + \mathcal{R} \right), \hspace{0.3in} \label{c2} \\
&&A_{-} E^{-1}  - A_{+} E = A_b, \mbox{   and  }  \hspace{0.2in} \label{c3}\\
&& A_{-}E^{-1}+A_{+} E=\frac{kA_b}{K} , \label{c4}
\end{eqnarray}
which determine $\mathcal{R}$, $A_{-}$, $A_{+}$, and $A_b$ given $\bar{B}_d$.  

When $\mathcal{R}$ is eliminated between Equations (\ref{c1}) and (\ref{c2}) and assuming $k/K<<1$,
\begin{equation}
A_{-}E+\frac{A_{+} }{E} = \frac{2\bar{B}_d}{K}.
\end{equation}
When $A_b$ is eliminated between Equations (\ref{c2}) and (\ref{c3}),
\begin{eqnarray}
A_+&=&-\frac{A_-}{E^2},  \mbox{   so   }\\
A(x) &=& A_{-}\left(e^{-Kx} +\frac{1}{E^2} e^{Kx} \right)   \mbox{   with   }\\
A_{-} &=& \frac{E^3}{E^4 - 1} \frac{2\bar{B}_d}{K}  \mbox{   and }  \hspace{0.2in}\\
(\vec{B}\cdot\hat{n})_\ell &=& \frac{2k}{K}\frac{E^4 +1}{E^4-1}\bar{B}_d \cos(ky) e^{t/\tau_g} . \label{norm-B} \\
(\vec{B}\cdot\hat{n})_r &=& \frac{4k}{K}\frac{E^2}{E^4-1}\bar{B}_d \cos(ky) e^{t/\tau_g}, \label{norm-Br}
\end{eqnarray}
where $(\vec{B}\cdot\hat{n})_\ell$ is evaluated at $x=-d_s/2$ and  $(\vec{B}\cdot\hat{n})_r$ is evaluated at $x=d_s/2$.

The surface current, $\bar{\mathcal{I}}\sin(ky) e^{t/\tau_g}\hat{z}$ is the current density integrated across the shell with $\vec{j} =j(x)\sin(ky) e^{t/\tau_g}\hat{z}$, where
\begin{eqnarray}
j(x) &=& - \frac{A(x)}{\eta\tau_g}\\
&=&-\frac{K^2}{\mu_0} A(x) \\
\bar{\mathcal{I}}&\equiv& \int_{-d_s/2}^{d_s/2} j(x)dx \\
&=& -\frac{K}{\mu_0}\big(A_- - A_+ \big)\left(E- E^{-1}\right)\\
&=&-\frac{K}{\mu_0} A_-\frac{E^2+1}{E^2} \frac{E^2-1}{E} \\
&=&-\frac{2\bar{B}_d}{\mu_0}.
\end{eqnarray}
The surface current has the same value as long as $k/K<<1$.

The jump in the normal component of the magnetic field is relative to its value on the left-hand side, at $x=-d_s/2$, is 
\begin{eqnarray}
\frac{\big[\vec{B}\cdot\hat{n}\big]}{(\vec{B}\cdot\hat{n})_\ell} &=& \frac{(\vec{B}\cdot\hat{n})_r-(\vec{B}\cdot\hat{n})_\ell}{(\vec{B}\cdot\hat{n})_\ell} \\
&=& - \frac{(E^2 -1)^2 }{E^4 + 1}. 
\end{eqnarray}
When the perturbation grows rapidly, $\tau_g<<\tau_s$, $E\rightarrow\infty$, the normal magnetic field on the right-hand side, $(\vec{B}\cdot\hat{n})_r$, is very small compared to the magnetic field on the left hand side, $(\vec{B}\cdot\hat{n})_\ell$. When the perturbation grows slowly, $\tau_g>>\tau_s$, $E^2\rightarrow 1 +Kd_s$ and the jump in the normal component divided by the normal component is $-(Kd_s)^2/2$, which is small since $Kd_s=\sqrt{\tau_s/\tau_g}<<1$.

The jump in the tangential magnetic field is
\begin{eqnarray}
\big[\vec{B}_{tan}\big] &=& -\Big[\frac{\partial A_z}{\partial x}\Big]\hat{y} \\
&=&-\Big[A'(x)\Big] \sin(ky) e^{t/\tau_g} \\
\Big[A'(x)\Big]&=&KA_{-}\Big[e^{-Kx} -\frac{1}{E^2} e^{Kx}\Big]  \\ 
&=&KA_{-}\left(E -\frac{1}{E^3}\right) \\
&=&2\bar{B}_d \\
&=&-\mu_0\bar{\mathcal{I}} \\
\big[\vec{B}_{tan}\big]&=& \mu_0\bar{\mathcal{I}} \sin(ky) e^{t/\tau_g}\hat{y}.
\end{eqnarray}
Independent of the $\tau_g/\tau_s$ ratio.  In a skin current model $\big[\vec{B}_{tan}\big]=-\mu_0\vec{\nabla}\kappa$, so $\kappa =(\mathcal{I}/k)\cos(ky) e^{t/\tau_g}$.

The magnetic field perpendicular to the driven side of the shell, at $x=-d_s/2$, is given in Equation (\ref{norm-B}).  In the slow growth limit $\tau_g>>\tau_s$, which is required for the validity of the current-potential approximation,
\begin{eqnarray}
\vec{B}\cdot\hat{n}&\rightarrow&2\frac{k\bar{B}_d}{K^2 d_s}\cos(ky) e^{t/\tau_g}  \mbox{   and   }  \\
\kappa &=& -d_s\frac{K^2}{k^2}\frac{\vec{B}\cdot\hat{n}}{\mu_0}
\end{eqnarray}
using the results for the jump in the tangential magnetic field. This equation for $\kappa$ is equivalent to Equation (\ref{cp-c}).

\vspace{0.1in}



\section{Shielding by magnetization \label{sec:magnetization} }

Linear magnetization means $\vec{B}=\mu \vec{H}$ with $\vec{\nabla}\times\vec{H}=\vec{j}_{free}$, which is the large-scale current density.  Thin layers of materials with $\mu/\mu_0>>1$ are described by a related but different mathematical form than thin shells of $\vec{j}_{free}$ current, which are the focus of this paper.  Unfortunately, no materials exist that exhibit linear magnetic properties with $\mu/\mu_0>>1$ except at low magnetic field strengths.  At low magnetic fields, mu-metal does with $\mu/\mu_0$ as large as $10^5$.  Nevertheless, the mathematical properties are interesting for comparison with those of a thin conducting or tiled shell.

The idealized problem is a thin shell of thickness $d_s$ of material in which $\mu/\mu_0>>1$.  As in a thin conducting or tiled shell, the effect on the external magnetic field is given by jump conditions across the shell on the normal, $B_\bot\equiv\hat{n}\cdot\vec{B}$, and the tangential magnetic field, $\vec{B}_{tan} \equiv \vec{B}-B_\bot\hat{n}$:
\begin{eqnarray}
\Big[ \vec{B}_{tan} \Big] &=& 0; \\
\Big[ B_\bot \Big] &=& - \vec{\nabla}\cdot (D_s \vec{B}_{tan}); \label{B-bot jump}\\
D_s &\equiv& \int_0^{d_s} \frac{\mu}{\mu_0} dr
\end{eqnarray}

The derivation of the jump conditions for idealized mu-metal begins with the expression for the current density, $\vec{j} =(\vec{\nabla}\ln\mu \times\vec{B})/\mu_0$, when $\vec{j}_{free}=0$.  The equation for the current density has two implications:  (1) The $\Big[ \vec{B}_{tan} \Big]$ jump condition.  (2) The tangential field is $\mu/\mu_0$ times stronger inside the mu-metal than it is outside.  The condition $\vec{\nabla}\cdot\vec{B}=0$ then implies the jump condition $\Big[ B_\bot \Big]$, Equation (\ref{B-bot jump}).

The two jump conditions give the magnetic field components on the two sides of a planar shell when a magnetic field of amplitude $B_n$ is driven from the negative $x$ side of the shell.
\begin{eqnarray}
&&\mbox{The field on the  } x<0 \mbox{    side is} \nonumber\\
&& \frac{B_x}{B_n} =\frac{1+kD_s}{1+\frac{1}{2}kD_s} \cos(ky), \hspace{0.1in}  \frac{B_y}{B_n} =\frac{  \sin(ky)}{1+\frac{1}{2}kD_s}. \nonumber\\
\nonumber\\
&&\mbox{The field on the } x>0 \mbox{    side is}  \nonumber\\
&&  \frac{B_x}{B_n} =\frac{ \cos(ky)}{1+\frac{1}{2}kD_s},  \hspace{0.2in}  \frac{B_y}{B_n} =\frac{ \sin(ky)}{1+\frac{1}{2}kD_s}.
\end{eqnarray}
This answer has an interesting relationship to the shielding produced by separated tiles, Equation (\ref{tiled field})


\end{document}